\documentclass{ws-ijmpa}
\usepackage{latexsym}
\usepackage{graphicx,epsfig}

\begin{document}
\markboth{Rakhi et al.}
{A Cosmological Model with Fermionic Field}

\catchline{}{}{}{}{}
\title{A COSMOLOGICAL MODEL WITH FERMIONIC FIELD}

\author{RAKHI R.\footnote{corresponding author.}}
\address{School of Pure and Applied Physics, Mahatma Gandhi University, Priyadarshini Hills Post,\\Kottayam, Kerala, India. PIN 686 560.
\\rakhir006@gmail.com}

\author{G. V. VIJAYAGOVINDAN \footnote{deceased}}
\address{School of Pure and Applied Physics, Mahatma Gandhi University, Priyadarshini Hills Post,\\Kottayam, Kerala, India. PIN 686 560.}

\author{INDULEKHA K.}
\address{School of Pure and Applied Physics, Mahatma Gandhi University, Priyadarshini Hills Post,\\Kottayam, Kerala, India. PIN 686 560.
\\kindulekha@gmail.com}

\maketitle

\begin{history}
\received{}
\revised{}
\end{history}

\begin{abstract}
In this work, a cosmological model inspired by string/M- theory with fermionic field is taken into consideration. Here it is investigated whether the introduction of a non-Dirac fermionic field--characterized by an \textit{interaction term}--affects the cosmological evolution.  The self-interaction potential is considered as a combination of the scalar and pseudo-scalar invariants. It is observed that the fermionic field under consideration behaves like an inflation field for the early Universe and later on, as a dark energy field. The late time acceleration becomes more prominent by the addition of the interaction term. There is a slight decrease for the inflation peak as well as for the energy density. We see that addition of higher order terms to the fermionic part of Lagrangian does not significantly change either the inflation or the late time acceleration behavior.
\end{abstract}

\ccode{PACS numbers: 98.80.-k, 98.80.Cq, 98.80.Qc}

\section{Introduction}
In cosmology, the investigation for the constituents responsible for the accelerated periods in the evolution of the universe is of great interest. The mysterious dark energy has been proposed as a cause for the late time dynamics of the current accelerated phase of the universe. Recently several approaches were made to explain accelerated expansion by choosing fermionic fields as the gravitational sources of energy.

Ribas et.al \cite{ribas}, in 2005, investigated whether fermionic field, with a self-interacting potential that depends on scalar and pseudo-scalar invariants, could be responsible for accelerated periods during the evolution of the Universe, where a matter field
would answer for the post inflation decelerated period. They have shown that the fermionic field behaves like an inflation field for the early Universe and later on,
as a dark energy field, whereas the matter field was created by an irreversible process connected with a non-equilibrium pressure. They observed that for the old decelerated phase of the universe the fermionic field plays the role of dark energy and drives the universe into an accelerated regime. But it is extremely difficult to present a completely satisfactory theory of dark energy. Hence cosmologists now look to string/M-theory to formulate a successful dark energy theory because string theory could provide some insight on the peculiar initial conditions required by standard big bang.  
Nojiri et.al \cite{nojiri} proposed the Gauss-Bonnet dark energy model, inspired by string/M-theory where standard gravity with scalar contains additional scalar-dependent coupling with Gauss-Bonnet invariant. Their study indicates that current acceleration may be significantly influenced by string effects. 

In the present work, we propose a cosmological model with fermionic field. Noticing that string/M-theory requires additional higher order terms, we introduce powers other than those that would emerge from the potential of Ribas et.al. To do so in a systematic way, a  term proportional to the square of the fermionic Lagrangian density, which we call as an \textit{interaction term}, is added to the total Lagrangian. It may be noted that the fermion condensate expected from super symmetry also calls for higher order terms.  In the present paper our aim is to investigate how the non-linear interaction term affects the cosmological evolution. 

\section{String Theory Inspired Cosmological Model with Fermionic Field}

The action for this model reads
\begin{equation}
\label{eq1}
S=\int {\sqrt {-g} } d^4x\left( {L_g +L_m +a_1 L_D +\frac{a_2}{M_{pl}^4} L_D^2} \right)
\end{equation}
where $L_g =R/2$, with $R$ denoting the curvature scalar, $L_g$ is the gravitational Einstein Lagrangian density, $L_m $ is the Lagrangian of the matter field and $a_1 $ and $a_2 $ are coupling constants. We are working under the assumption that higher powers of $L_D$ grow less important. The $L_D^2$ term is suppressed relative to the main terms by the fourth power of the Planck mass, $M_{pl}$ . (The $L_D^3$ term would have eighth power of $M_{pl}$ suppressing it relative to the main term etc). The values of coupling constants $a_1$ and $a_2$ cannot be bigger than unity for the same reason; otherwise the assumption of ordering would be wrong. We have $M_{pl}^{-2}=8 \pi G$  . Later, this is set to be equal to unity.  Finally, the Dirac Lagrangian density $L_D$ of the fermionic field \cite{ribas},\cite{green} for a fermionic mass $m$ is given by
\begin{equation}
\label{eq2}
L_D =\frac{i}{2}\left[ {\overline \Psi \Gamma ^\mu D_\mu \Psi -\left( {D_\mu
\overline \Psi } \right)\Gamma ^\mu \Psi } \right]-m\left\langle {\overline
\Psi \Psi } \right\rangle -V 
\end{equation}
In Eq. (\ref{eq2}), $V=V(\Psi,\overline \Psi )$ describes the potential density of self-interaction between fermions and $\overline \Psi =\Psi ^\dag \gamma ^0$ denotes the adjoint spinor field. Moreover, the connection between general relativity and Dirac equation is done via the tetrad formalism and the components of the tetrad play the role of gravitational degrees of freedom. That is, $\Gamma ^\mu =e_a^\mu \gamma ^a$ are the generalized Dirac-Pauli matrices, where $e_a^\mu $ denote the tetrad or \textit{vierbein}\cite{birrell}. The covariant derivatives in Eq. (\ref{eq2}) are given by,
\[
D_\mu \Psi =\partial _\mu \Psi -\Omega _\mu \Psi
\]
\begin{equation}
\label{eq3}
D_\mu \overline \Psi =\partial _\mu \overline \Psi +\overline \Psi \Omega _\mu
\end{equation}
with the spin connection,
\begin{equation}
\label{eq4}
\Omega _\mu =-\frac{1}{4}g_{\rho \sigma } [\Gamma _{\mu \delta }^\rho
-e_b^\rho \partial _\mu e_\delta ^b ]\Gamma ^\sigma \Gamma ^\delta
\end{equation}
where $\Gamma _{\mu \delta }^\rho $ is the Christoffel symbol.

\subsection{Basic Equations}
Through Euler-Lagrange equations, from Eqs. (\ref{eq1}) and (\ref{eq2}), we can obtain the equations of motion for the spinor field as
\begin{equation}
\label{eq5}
\left(a_1+2 \frac{a_2}{M_{pl}^4} L_D\right) dL_D/d\overline \psi=0
\end{equation}which requires that either $\left(a_1+2 \frac{a_2}{M_{pl}^4} L_D\right)=0$
  or   $dL_D/d\overline \psi=0$.

The first case, i.e. $\left(a_1+2 \frac{a_2}{M_{pl}^4} L_D\right)=0$ , gives $L_D=-\frac{a_1}{2\left( a_2/M_{pl}^4\right)}$. This appears to be an unphysical constraint, and is not explored here further.

We take $dL_D/d\overline \psi=0$ as our ansatz. I.e.
\begin{equation}
\label{eq6}
i\Gamma ^\mu D_\mu \Psi -m\Psi - \frac{dV}{d\overline \Psi }=0
\end{equation}
The equation of motion is exactly the same as in the case where no $L_D^2$ term
is present. However we see that the $L_D^2$ term modifies the source terms for gravity
in a significant way. We also note that any series of arbitrary powers $L_D^n$ present
in the original Lagrangian would lead to $\delta L_D/\delta \psi =0$ for the fermionic equation of motion, because the other equation arising would be unphysical. So, with the above ansatz,  our analysis can be thought of as treating the case of a general form for the fermionic contribution to the stress tensor, where we retain terms only upto the next to leading order contribution in an expansion of the stress tensor in the derivatives of $\psi$.

Similarly, we can obtain the Dirac equations for the adjoint coupled to the
gravitational field.

The variation of the action Eq.(\ref{eq1}) with respect to the tetrad gives the
Einstein's field equations as
\begin{equation}
\label{eq7}
R_{\mu \nu } -\frac{1}{2}g_{\mu \nu } R=-T_{\mu \nu }
\end{equation}
where $T_{\mu \nu } =T_f^{\mu \nu } +T_m^{\mu \nu } $, with $T_f^{\mu \nu } $ being the energy-momentum tensor of the fermionic field and $T_m^{\mu \nu} $, that of the matter field and $T_{\mu \nu }=-\frac{2}{\sqrt{-g}}\frac{\delta S}{\delta g^{\mu \nu}}$ . The symmetric form of the energy-momentum tensor of the fermionic field \cite{green,birrell} which follows from Eq. (\ref{eq1}) gives
\[
T_f^{\mu \nu } =\frac{i}{4}\left\{ {\overline \Psi \Gamma ^\mu D^\nu \Psi
+\overline \Psi \Gamma ^\nu D^\mu \Psi -D^\nu \overline \Psi \Gamma ^\mu
\Psi -D^\mu \overline \Psi \Gamma ^\nu \Psi } \right\}+\]
\begin{eqnarray}
\nonumber \hspace{90pt} \frac{i}{2}L_D
\{\overline \Psi \Gamma ^\mu D^\nu \Psi +\overline \Psi \Gamma ^\nu D^\mu
\Psi -D^\nu \overline \Psi \Gamma ^\mu \Psi -D^\mu \overline \Psi \Gamma ^\nu \Psi
\}-\end{eqnarray}
\begin{eqnarray}
\label{eq8}
g^{\mu \nu }a_1 L_D-g^{\mu \nu } \frac{a_2}{M_{pl}^4} L_D ^2
\end{eqnarray}
where the Dirac matrices have the form $\Gamma ^0=\gamma ^0,\Gamma
^i=\frac{1}{a(t)}\gamma ^i,\Gamma ^5=\gamma ^5$ and the spin connection
components are
\begin{equation}
\label{eq10}
\Omega _0 =0,\Omega _i =\frac{1}{2}\mathop a\limits^. (t)\gamma ^i\gamma ^0
\end{equation}
Based upon the cosmological principle, which states that our universe is homogenous and isotropic, we use the Robertson-Walker metric to describe our Universe:
\begin{equation}
\label{eq11}
ds^2=dt^2-a(t)^2\left( {dx^2+dy^2+dz^2} \right)
\end{equation}
The interactions between the constituents are modeled through the presence of a non-equilibrium pressure term ($\varpi$)  in the source's energy-momentum tensor. Therefore, for an isotropic, homogenous Universe, the total energy-momentum tensor- which is composed by fermionic and matter fields- can be written as
\begin{equation}
\label{eq10c}
T_\mu^\nu=diag(\rho,-p-\varpi,-p-\varpi,-p-\varpi)
\end{equation}
where $\rho=\rho_f+\rho_m$ and $p=p_f+p_m$ and the quantity $\varpi$ refers to a non-equilibrium pressure which is related to dissipative processes during the evolution of the universe and represents an irreversible process of energy transfer between the matter and the gravitational field.
By the use of Eq.(\ref{eq10c}), the components of the energy-momentum tensor of the fermionic can be identified as
\begin{equation}
\label{eq10d}
(T_f)_\mu^\nu=diag(\rho_f,-p_f,-p_f,-p_f)
\end{equation}
Now the non-vanishing components of the energy-momentum tensor for the fermionic field following from Eqs. (\ref{eq1}), (\ref{eq7}) and (\ref{eq11}) can be written as, 
\begin{equation}
\label{eq10ab}
\left( {T_f } \right)_0^0 =a_1 \left( {m\left\langle {\overline \Psi \Psi }
\right\rangle +V } \right)-\frac{a_2}{M_{pl}^4} \left( {m\left\langle {\overline \Psi
\Psi } \right\rangle +V } \right)^2+\frac{a_2}{M_{pl}^4} \left( {m\left\langle
{\overline \Psi \Psi } \right\rangle +\frac{\overline \Psi
}{2}\frac{dV}{d\overline \Psi } +\frac{\Psi }{2}\frac{dV}{d\Psi
} } \right)^2
\end{equation}
\begin{equation}
\label{eq10a}
(T_f )_1^1 =(T_f )_2^2 =(T_f )_3^3 =a_1  \left[ {V-\frac{\overline
\Psi }{2}\frac{dV}{d\overline \Psi }-\frac{dV}{d\Psi }\frac{\Psi }{2}}
\right]-\frac{a_2}{M_{pl}^4} \left[ {V-\frac{\overline \Psi }{2}\frac{dV}{d\overline
\Psi }-\frac{dV}{d\Psi }\frac{\Psi }{2}} \right]^2
\end{equation}
Now, the self-interaction potential takes the form,
\begin{equation}
\label{eq13}
V=\lambda \left[ {\beta _1 \left( {\left\langle {\overline \Psi \Psi }
\right\rangle } \right)^2+\beta _2 \left( {i\left\langle {\overline \psi
\gamma ^5\Psi } \right\rangle } \right)^2} \right]^n
\end{equation}
where $\lambda $ is the coupling constant and $n$ is a constant exponent. We
consider $V$ as a combination of the scalar and pseudo-scalar invariants.We assume both the scalar and pseudo-scalar densities to play equally important roles and assume the coefficients  $\beta_1$ and $\beta_2$ to be both O (1). For simplicity we consider the values to be unity.
That is, $\beta _1 =\beta _2 =1$.

Combining  Eqs. (\ref{eq10d}), (\ref{eq10ab}), (\ref{eq10a}) and (\ref{eq13}) we solve for the energy density and
pressure of the fermionic field as
\begin{equation}
\label{eq14}
\rho _f =a_1 \left( {m\left\langle {\overline \Psi \Psi } \right\rangle
+V } \right)-\frac{a_2}{M_{pl}^4} \left( {m\left\langle {\overline \Psi \Psi }
\right\rangle +V } \right)^2+\frac{a_2}{M_{pl}^4} \left( {m\left\langle {\overline \Psi
\Psi } \right\rangle +2nV } \right)^2
\end{equation}
\begin{equation}
\label{eq15}
p_f =a_1 \left( {2n-1} \right)V +\frac{a_2}{M_{pl}^4} \left[ {\left( {2n-1}
\right)V } \right]^2
\end{equation}
respectively. We infer from Eq.(\ref{eq15}) that the fermions could be classified according to the value of the exponent $n$. Indeed, for $n\ge 1/2$ , the fermions represent a matter field with positive pressure ($n > 1/2$) or a pressure-less fluid ($n = 1/2$), whereas for $n < 1/2$ the pressure of the fermions is negative and they could represent either the inflation or the dark energy. For the massive case, we shall deal with only the case where the fermionic field behaves as inflation or dark energy, i.e., the case where $n < 1/2$. \cite{ribas}

The Friedman equations are
\begin{equation}
\label{eq16}
3H^2=\frac{8\pi G}{c^2}T_0^0=\frac{8\pi G}{c^2}\rho =\frac{8\pi G}{c^2}(\rho _f +\rho _m)
\end{equation}
\begin{equation}
\label{eq16a}
2\frac{\mathop a\limits^{..} }{a}=-\frac{8\pi G}{3 c^2}\left[T_0^0+3 T_1^1\right] 
\end{equation}
or 
\begin{equation}
\label{eq16b}
2\frac{\mathop a\limits^{..} }{a}=-\frac{8\pi G}{3 c^2}\left[\rho_f+\rho_m+3 p_f+3 p_m+3 \varpi\right]
\end{equation}

Using Eqs. (\ref{eq16}) and (\ref{eq16b}), the acceleration equation becomes
\begin{equation}
\label{eq17}
\frac{\mathop a\limits^{..} }{a}=-\frac{8\pi G}{6 c^2}\left[\rho_f+\rho_m+3 p_f+3 p_m+3 \varpi\right]
\end{equation}
The covariant differentiation of Einstein field equations, using the Bianchi identities, leads to the conservation law of the total energy-momentum tensor: i.e $T^{\mu \nu}_{;\nu}=0 $.

I.e.
\begin{equation}
\label{eq18}
\mathop {\rho}\limits^. +3H\left( {\rho +p +\varpi } \right)=0
\end{equation}
where $H=\frac{\mathop {a(t)}\limits^. }{a}$ is the Hubble's parameter.
The conservation law for the energy density of the fermionic field and
matter field can thus be written as
\begin{equation}
\label{eq18a}
\mathop {\rho _f }\limits^. +3H\left( {\rho _f +p_f } \right)=0
\end{equation}
\begin{equation}
\label{eq19}
\mathop {\rho _m }\limits^. +3H\left( {\rho _m +p_m +\varpi } \right)=0
\end{equation}

Considering the fermionic field to be a function of time alone, the Dirac
equations Eq.(\ref{eq6}) becomes
\begin{equation}
\label{eq20}
\mathop \Psi \limits^. +\frac{3}{2}H\Psi +im\gamma ^0\Psi +i\gamma ^0
\frac{dV}{d\overline \Psi }=0
\end{equation}
Similarly, the Dirac equation for the adjoint spinor field coupled to the
gravitational field becomes
\begin{equation}
\label{eq21}
\mathop {\overline \Psi }\limits^. +\frac{3}{2}H\overline \Psi -im\overline
\Psi \gamma ^0-i \frac{dV}{d\overline \Psi }\gamma ^0=0
\end{equation}

\subsection{Field Equations}

The system of field equations that help us to find the cosmological
solutions is:

\subsubsection{The Acceleration Equation }
The acceleration equation for the cosmological model under consideration can
be obtained from equations (\ref{eq14}), (\ref{eq15}) and (\ref{eq17}).We have chosen the units so that $\hbar=c=1$. Also we consider the
pressure of the matter field as $p_m =w_m \rho _m $ with $0\le w_m \le 1$,
which is a barotropic equation of state. We again consider a vanishing
energy density of the matter field $\left( {\rho _m } \right)$ and a
vanishing non-equilibrium pressure $\left( \varpi \right)$at t=0.
Incorporating all these ideas, the normalized acceleration equation for the present
model becomes
\begin{equation}
\label{eq22}
\frac{\mathop a\limits^{..} }{a}=-\frac{8\pi G}{6}\left(\rho_f+\rho_m+3 p_f+3 p_m+3 \varpi\right)
\end{equation}
where \newline
$
\rho _f =a_1 \left( {m\left\langle {\overline \Psi \Psi } \right\rangle
+V } \right)-\frac{a_2}{M_{pl}^4} \left( {m\left\langle {\overline \Psi \Psi }
\right\rangle +V } \right)^2+\frac{a_2}{M_{pl}^4} \left( {m\left\langle {\overline \Psi
\Psi } \right\rangle +2nV } \right)^2
$\newline
$
p_f =a_1 \left( {2n-1} \right)V +a_2 \left[ {\left( {2n-1}
\right)V } \right]^2
$
\newline\newline
Putting $8 \pi G=1$ , Eq. (\ref{eq22}) will become
\begin{equation}
\label{eq22b}
\frac{\mathop a\limits^{..} }{a}=-\frac{1}{6}\left(\rho_f+\rho_m+3 p_f+3 p_m+3 \varpi\right)
\end{equation}
with\newline $
\rho _f =a_1 \left( {m\left\langle {\overline \Psi \Psi } \right\rangle
+V } \right)-a_2 \left( {m\left\langle {\overline \Psi \Psi }
\right\rangle +V } \right)^2+a_2 \left( {m\left\langle {\overline \Psi
\Psi } \right\rangle +2nV } \right)^2
$
\newline
$
p_f =a_1 \left( {2n-1} \right)V +a_2 \left[ {\left( {2n-1}
\right)V } \right]^2
$
\newline\newline
For further calculations we use this result.
\subsubsection{Evolution equation for the energy density of matter field}
The evolution equation for $\rho _m $ is given by
\begin{equation}
\label{eq24}
\mathop {\rho _m }\limits^. +3H(\rho _m +w_m \rho _m +\varpi )=0
\end{equation}

\subsubsection{Evolution equation for the non-equilibrium pressure}

The linearised form of the evolution equation for non-equilibrium pressure\cite{kremer1,kremer2,kremer3} reads
\begin{equation}
\label{eq25}
\tau \mathop \varpi \limits^. +\varpi +3\eta H=0
\end{equation}
\newline
where the coefficient of bulk viscosity $\eta$  and the characteristic time $\tau$  are assumed to be related with the energy density $\rho$ by $\eta=\alpha \rho$ and $\tau=\frac{\eta}{\rho}$ , where $\alpha$ is a constant.
\subsubsection{The Dirac Equation }
In terms of the spinor components,$\Psi =\left( {\Psi _1 ,\Psi _2 ,\Psi _3
,\Psi _4 } \right)^T$, the Dirac Eq. (\ref{eq20}) can be written as
\begin{equation}
\label{eq26}
\frac{d}{dt}\left( {\begin{array}{l}
 \Psi _1 \\
 \Psi _2 \\
 \Psi _3 \\
 \Psi _4 \\
 \end{array}} \right)+\frac{3}{2}H\left( {\begin{array}{l}
 \Psi _1 \\
 \Psi _2 \\
 \Psi _3 \\
 \Psi _4 \\
 \end{array}} \right)+im\left( {\begin{array}{l}
 \Psi _1 \\
 \Psi _2 \\
 -\Psi _3 \\
 -\Psi _4 \\
 \end{array}} \right)-2i\left( {\Psi
_1^\dag \Psi _1 +\Psi _2^\dag \Psi _2 -\Psi _3^\dag \Psi _3 -\Psi _4^\dag
\Psi _4 } \right)\left( {\begin{array}{l}
 \Psi _1 \\
 \Psi _2 \\
 -\Psi _3 \\
 -\Psi _4 \\
 \end{array}} \right)\frac{dV}{d\left( {\overline \Psi \Psi } \right)^2}
\quad
\end{equation}
$-2i\left( {\Psi _3^\dag \Psi _1 +\Psi
_4^\dag \Psi _2 -\Psi _1^\dag \Psi _3 -\Psi _2^\dag \Psi _4 } \right)\left(
{\begin{array}{l}
 \Psi _3 \\
 \Psi _4 \\
 -\Psi _1 \\
 -\Psi _2 \\
 \end{array}} \right)\frac{dV}{d\left( {\overline \Psi \gamma ^5\Psi }
\right)^2}=0
$

\subsection{Cosmological Solutions}

In order to obtain numerical solutions of the coupled system of Eqs. (\ref{eq22})-
(\ref{eq26}), we have to specify the parameters $\lambda ,\beta _1 ,\beta _2
,n,m,\alpha ,w_m $ and $a_1 $. These parameters take the following values:

$\lambda=0.1 ,\beta _1=\beta _2=1,n=0.25,m=0.01,\alpha=1,w_m=1/3$ and $a_1=1$.

\subsubsection{Dependence of the evolution on the coefficient of the interaction term $a_2$}
The initial conditions, we have chosen for t=0 (by adjusting clocks) are: 
\newline
$a\left( 0\right)=1,\Psi _1 \left( 0 \right)=0.1i,\Psi _2 \left( 0 \right)=1,\Psi _3
\left( 0 \right)=0.3,\Psi _4 \left( 0 \right)=i,\rho _m \left( 0
\right)=0.05,\varpi \left( 0 \right)=0.$ The value of $\rho _m \left( 0
\right)$ is so chosen that the observable normalized matter density is 5\%.  The last initial condition corresponds to a vanishing non-equilibrium pressure at t = 0. The conditions chosen here characterize qualitatively an initial proportion between the constituents in the corresponding era.

Graphs are plotted for various values of $a_{2}$.
The effect of $L_D ^2$ is now studied first by choosing $a_2 =0.5$ [Figs.1-2].
\newline
In Fig. 1, we plot the evolution of accleration versus time for $a_2 =0.5$ while Fig.3 shows the the evolution of accleration versus time for $a_2 =0$. In Fig. 2,  the evolution of energy densities of fermionic field and matter field is plotted.
\begin{figure}[htbp]
\centerline{\includegraphics[width=2.53in,height=1.5in]{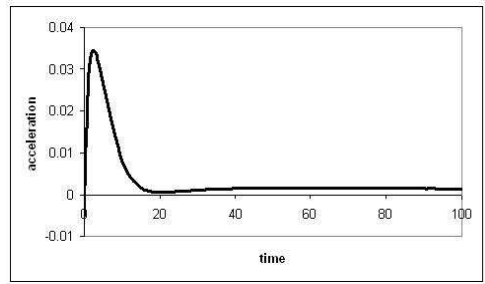}}
\label{fig1}
\caption{Acceleration vs. time $a_2 =0.5$}
\end{figure}
\begin{figure}[htbp]\centerline{\includegraphics[width=2.53in,height=1.50in]{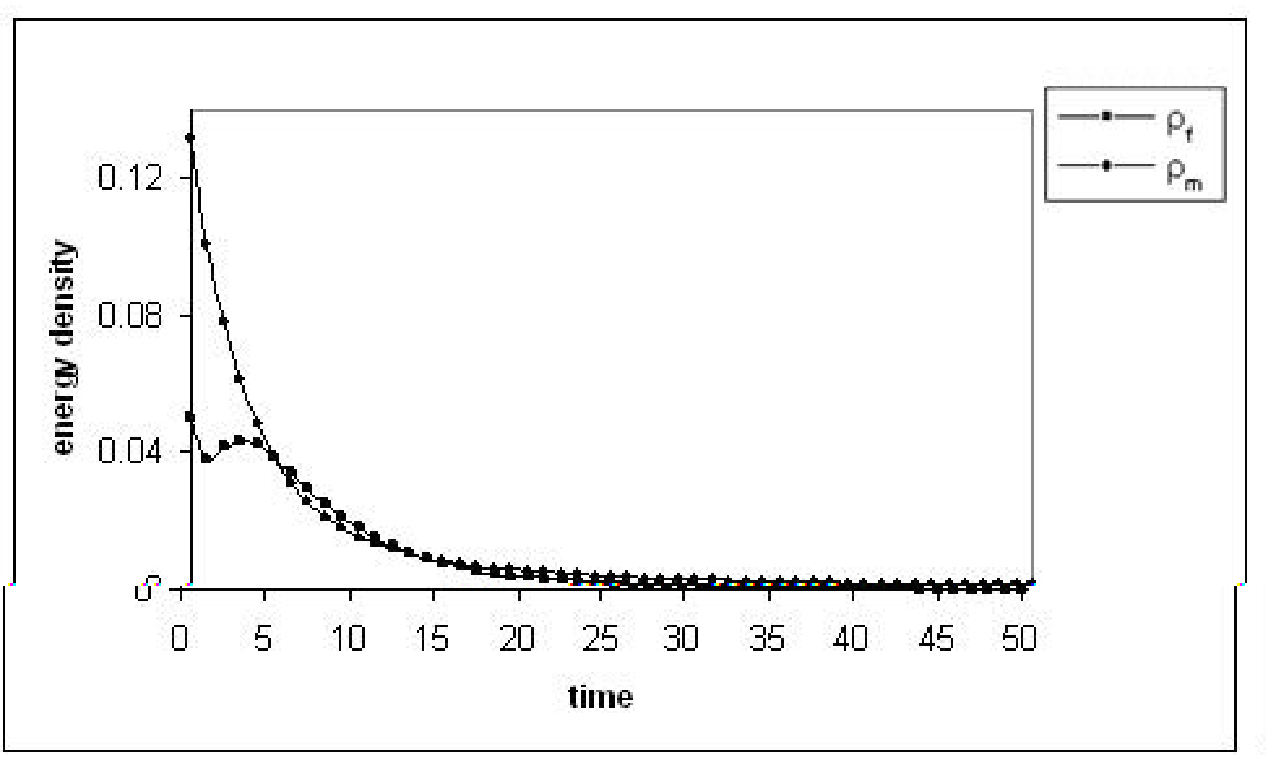}}
\label{fig2}
\caption{Energy density of fermionic $\rho_f$ and matter $\rho_m$ fields vs. time}
\end{figure}

\begin{figure}[hbtp]
\centerline{\includegraphics[width=2.53in,height=1.5in]{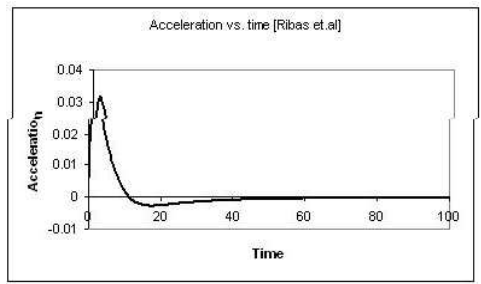}}
\label{fig3}
\caption{Acceleration vs. time for $a_2 =0$[Ribas et.al(2005)]}
\end{figure}

The effect of $L_D ^2$ for various values of $a_2$ is studied below by taking $a_2 =0.1, 0.2, 0.8$ respectively [Figs. 4-5].
\begin{figure}[h]
\centerline{\includegraphics[width=2.53in,height=1.5in]{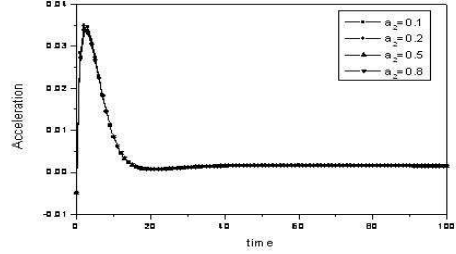}}
\label{fig4}
\caption{Acceleration vs. time for various $a_2$}
\end{figure}

\begin{figure}[htbp]
\centerline{\includegraphics[width=2.53in,height=1.5in]{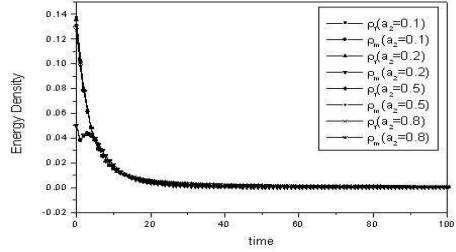}}
\label{fig5}
\caption{Energy density of fermionic $\rho_f$ and matter $\rho_m$ fields vs. time for various $a_2$}
\end{figure}

In Fig. (1) and Fig. (4), acceleration vs. time is plotted whereas in Fig. (2) and Fig. (5), it is shown the behavior of the energy densities of the fermionic $\rho_f$ and matter $\rho_m$ fields as a function of time . We infer from these figures that the proposed model behaves like an inflation field for the early Universe and later on, as a dark energy field. Here, the late time acceleration becomes more prominent by the addition of the interaction term. There is a slight decrease for the inflation peak. 

\subsubsection{Study of evolution for $\left(\bar{\psi}\psi/M_{pl}^3\right)<<1$ and $\left(\bar{\psi}\psi/M_{pl}^3\right)>>1$}
We have already mentioned that the power $n$ of the scalar and pseudo-scalar densities in Eq.\ref{eq13} is required to be less than 0.5 for the fermionic field to behave like a dark energy at late times. With the addition of the interaction term, $n$ will have to be less than 0.25 to avoid suppression of this term. Then if $\left(\bar{\psi}\psi/M_{pl}^3\right)<<1$ (i.e. $\bar{\psi}\psi<1$ in Eq.\ref{eq22b}), the relative contribution from the $V$ in $L_D^2$will be larger compared to the case when $\bar{\psi}\psi>1$. Thus the new  effects we obtain with the addition of the interaction term should be more prominent when scalar or pseudo-scalar density is less than one, and when the densities become greater than one, the results obtained with the addition of the interaction term should be closer to the behavior seen with $L_D$ alone.

In order to check this, we plot the acceleration versus time, when scalar or pseudo-scalar density (at t=0) is greater than one [Fig. 6], and less than one [Fig.7] , for $a_2=0$ and for  $a_2=0.5$ in each case.
\newline\newline The initial conditions, we have chosen for $\bar{\psi}\psi<1$ at t =0 are:
$a\left( 0
\right)=1,\Psi _1 \left( 0 \right)=0.1i,\Psi _2 \left( 0 \right)=1,\Psi _3
\left( 0 \right)=0.3,\Psi _4 \left( 0 \right)=i,\rho _m \left( 0
\right)=0.05,\varpi \left( 0 \right)=0.$
\newline
whereas the initial conditions chosen for $\bar{\psi}\psi>1$ at t =0 are:
$a\left( 0\right)=1,\Psi _1 \left( 0 \right)=0.01i,\Psi _2 \left( 0 \right)=0.01i,\Psi _3
\left( 0 \right)=1,\Psi _4 \left( 0 \right)=i,\rho _m \left( 0
\right)=0.05,\varpi \left( 0 \right)=0.$

\begin{figure}[htbp]
\centerline{\includegraphics[width=2.83in,height=1.8in]{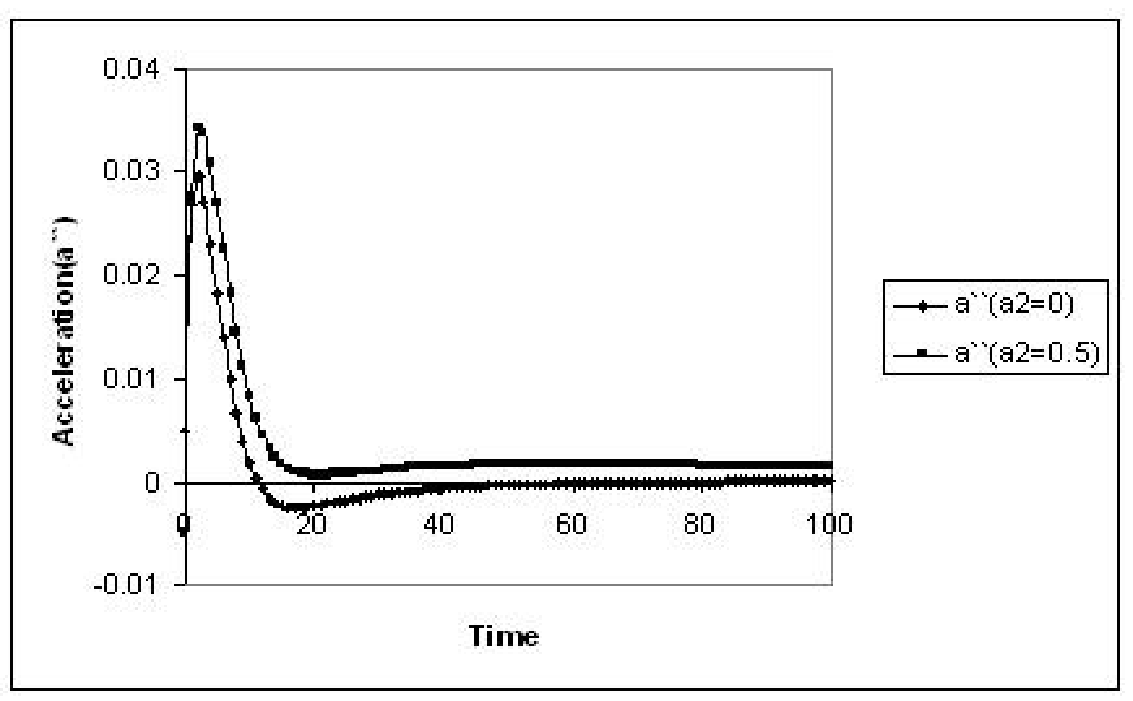}}
\label{fig6}
\caption{Acceleration Vs time for $\bar{\psi}\psi<1$}

\centerline{\includegraphics[width=2.93in,height=1.86in]{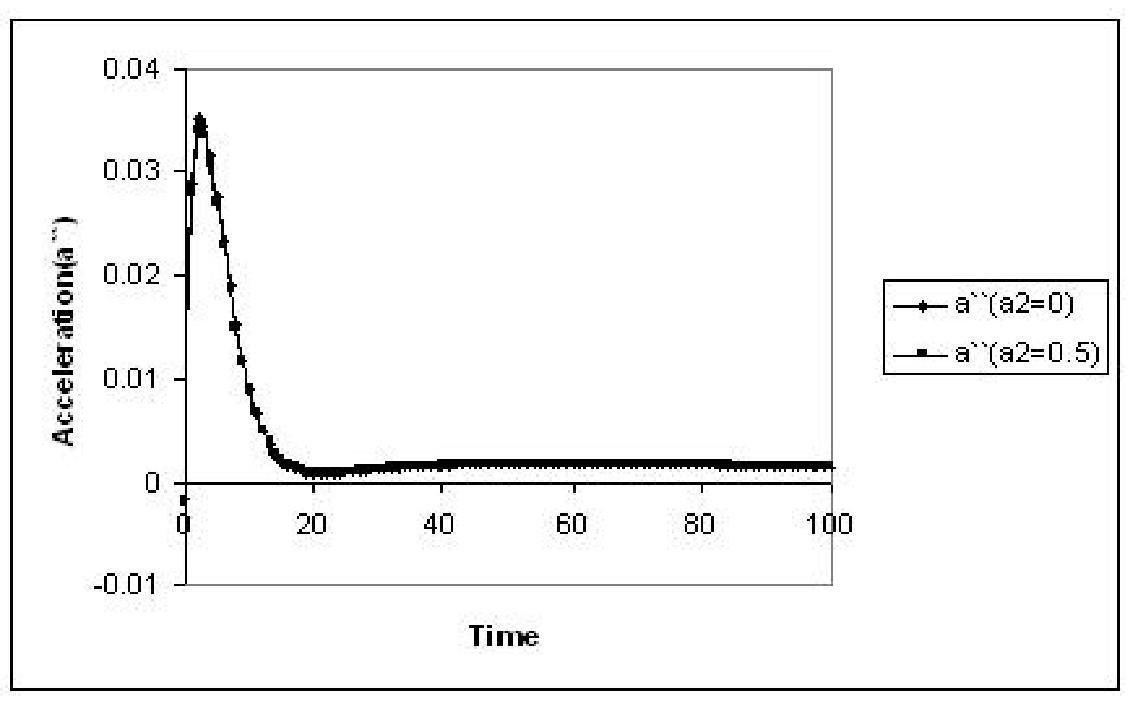}}
\label{fig7}
\caption{Acceleration Vs time for $\bar{\psi}\psi>1$}
\end{figure}
From the figures, we notice that the change in the accelerating behavior produced by the addition of the interaction term is more prominent, when scalar or pseudo-scalar density (at t=0) is less than one. In comparison we note that the difference is quite small when scalar or pseudo-scalar density (at t=0) is greater than one, as expected.

\subsection{Final Remarks and Conclusions}
In the present work, we have considered a cosmological model inspired by string/M- theory with fermionic field. Here we have assumed that there exists a field with a kinetic term like $L_D^2$ which we call as an \textit{interaction term} and this field corresponds to dark energy which contributes to expansion/inflation etc. The self-interaction potential is considered as a combination of the scalar and pseudo-scalar invariants.

We have investigated the contribution to the cosmological evolution of a non-Dirac fermionic field--characterized by an interaction term $L_D^2$. Early universe theories require the presence of higher order terms. To study these in a systematic way, we have included the $L_D^2$ term in the total Lagrangian and checked the effects of this non-linearity. We see that the model behaves like an inflation field for the early Universe and later on, as a dark energy field, similar to what Ribas et.al. obtained [Fig.3] with the addition of $L_D$ only.
Our results show that the higher order non-linear terms do not significantly change the overall behavior, obtained with $L_D$ alone; except that the late time acceleration is even more prominent. Also it is significant that the change in the accelerating behavior produced by the addition of the interaction term is more prominent, when scalar or pseudo-scalar density (at t=0) is less than one, whereas the difference is quite small when scalar or pseudo-scalar density (at t=0) is greater than one. So we conclude that fermionic terms to all orders in the Lagrangian lead to inflation at early times and acceleration at late times for the universe. 
Our model can generate standard acceleration similar to Dirac behavior as well as a non-Dirac behavior by simply adjusting the magnitude of the coupling constant $a_2$ . 

\section*{Acknowledgement}

The authors gratefully acknowledge fruitful discussions with Urjit A. Yajnik and Sasmitha Misra and their keen interest and help in the preparation of this manuscript.

\end{document}